# Spin waves excited by hard x-ray transient gratings

Peter R. Miedaner†,[1] Alexei A. Maznev†,[1] Mykola Biednov,[2] Marwan Deb,[3] Carles Serrat,[4] Nadia Berndt,[1] Pietro Carrara,[5] Cristian Soncini,[6] Marta Brioschi,[7,8] Daniele Ronchetti,[9,10,11] Andrei Benediktovitch,[11] Danny Fainozzi,[12] Nupur Khatu,[2,6,13] Eugenio Ferrari,[11] Joan Vila-Comamala,[14] Peter Zalden,[2] Wojciech Gawelda,[15,16,17] Martin Knoll,[2] Paul Frankenberger,[2] Ludmila Leroy,[14] Talgat Mamyrbayev,[14] Grigory Smolentsev,[14] Simon Gerber,[14] Alessandro Gessini,[6] Filippo Bencivenga,[6] Riccardo Cucini,[8] Giorgio Rossi,[7,8] Riccardo Mincigrucci,[6] Ettore Paltanin,[6] Majed Chergui,[6,18] Claudio Masciovecchio,[6] Stefano Bonetti,[13] Yohei Uemura,[2] Xinchao Huang,[2] Han Xu,[2] Frederico Alves Lima,[2] Fernando Ardana-Lamas,[2] Anders Madsen,[2] Renato Torre,[19] Luis Bañares,[21] Jakub Szlachetko,[22] Wojciech Blachucki,[23] Matias Bargheer,[24] Thomas Feurer,[2] Robin Y. Engel,[11] Martin Beye,[11,25] Christian David,[14] Urs Staub,[14] Andrea Cannizzo,[12] Cristopher Milne,[2] Keith A. Nelson,[1] Cristian Svetina†[2,16]

[1] Massachusetts Institute of Technology, Cambridge, MA, USA
[2] European XFEL GmbH, Schenefeld, Germany
[3] Institut des Molécules et Matériaux du Mans, Le Mans Université, Le Mans, France
[4] Polytechnic University of Catalonia, Barcelona, Spain
[5] Sorbonne Université, CNRS, Institut des NanoSciences de Paris, INSP, F-75005 Paris, France
[6] Elettra-Sincrotrone Trieste, Trieste, Italy.
[7] Dipartimento di Fisica, Università degli Studi di Milano, Milano, Italy.
[8] CNR-Istituto Officina dei Materiali, Trieste, Italy
[9] Department of Physics, Universität Hamburg, Hamburg 22761, Germany
[10] Max Planck School of Photonics, Friedrich-Schiller University of Jena, Jena 07745, Germany
[11] Deutsches Elektronen-Synchrotron DESY, Notkestr. 85, 22607 Hamburg, Germany
[12] University of Bern, Bern, Switzerland
[13] Department of Molecular Sciences and Nanosystems, Ca' Foscari University of Venice, 30172 Venice, Italy
[14] PSI Center for Photon Science, Paul Scherrer Institute, Villigen, Switzerland
[15] Universidad Autónoma de Madrid, Madrid, Spain
[16] Madrid Institute for Advanced Studies, IMDEA Nanociencia, Cantoblanco, 28049 Madrid, Spain
[17] Faculty of Physics, Adam Mickiewicz University, Poznań, Poland
[18] Lausanne Centre for Ultrafast Science, ISIC-FSB, École Polytechnique Fédérale de Lausanne, Switzerland
[19] European Laboratory for Non-Linear Spectroscopy and Dipartimento di Fisica ed Astronomia, Università di Firenze, Florence, Italy
[21] Universidad Complutense de Madrid, Madrid, Spain
[22] Solaris National Synchrotron Radiation Centre, Kraków, Poland
[23] Institute of Nuclear Physics, Polish Academy of Sciences, Kraków, Poland
[24] University of Potsdam, Potsdam, Germany
[25] University of Stockholm, Stockholm, Sweden
† miedaner@mit.edu, alexei.maznev@gmail.com, cristian.svetina@imdea.org

March 2025

Recent progress in ultrafast x-ray sources helped establish x-rays as an important tool for probing lattice and magnetic dynamics initiated by femtosecond optical pulses. Here, we explore the potential of ultrashort hard x-ray pulses for *driving* magnetic dynamics. We use a transient grating technique in which a spatially periodic x-ray excitation pattern gives rise to material excitations at a well-defined wave vector, whose dynamics are monitored via diffraction of an optical probe pulse. The excitation of a ferrimagnetic gadolinium bismuth iron garnet film placed in an external tilted magnetic field by x-rays at the Gd $L_3$ edge results in both magnetic and non-magnetic transient gratings whose contributions to the diffracted signal are separated by polarization analysis. We observe the magnetization precession at both longitudinal acoustic and spin wave frequencies. An analysis with the Landau-Lifshitz-Gilbert equation indicates that the magnetization precession is driven by strain resulting from thermal expansion induced by absorbed x-rays. The results establish x-ray transient gratings as a tool for driving coherent phonons and magnons, with the potential of accessing wave vectors across the entire Brillouin zone.

The remarkable advances in high-brightness ultrafast x-ray sources have opened many opportunities for studying ultrafast and nanoscale dynamics in condensed matter systems [1,2]. In particular, x-rays have been established as useful probes of magnetic dynamics excited by ultrashort optical pulses [3-6]. However, the use of hard x-rays for driving magnetic dynamics has remained largely unexplored. Very recently, ultrafast demagnetization [7], magnetization switching [8,9], and coherent magnon excitation [10] by extreme ultraviolet radiation (EUV) sourced from a free-electron laser (FEL) with wavelengths of 8 – 40 nm were demonstrated in transient grating (TG) experiments. In these experiments, a pair of pump pulses were crossed to generate a spatially periodic excitation pattern, and the dynamical response was monitored via diffraction of a time-delayed EUV probe pulse. The key advantage of using short wavelengths is the ability to produce small TG periods, enabling the excitation of transient nanoscale magnetic textures and generation of coherent magnons with nanoscale wavelengths. TG measurements have also recently been demonstrated at hard x-ray wavelengths on nonmagnetic systems [11,12]. Although the use of an optical probe in Ref. [11] limited the TG period range, hard x-rays can potentially yield much shorter TG periods than currently possible with EUV excitation. Other advantages of hard x-rays are the ability to study bulk materials due to the long penetration lengths of x-rays compared to EUV radiation and the prospect of achieving element specificity by using resonant excitation at x-ray absorption edges and resonant nonlinear wave mixing [13,14].

In this report, we explore the potential of the x-ray transient grating (XTG) technique for driving magnetic dynamics in gadolinium bismuth iron garnet (GdBiIG), an insulating ferrimagnet with attractive properties for magnetooptical applications [15]. Following Refs [11,16], we use Talbot imaging of a phase grating to form a periodic x-ray intensity pattern to pump the sample, and the ensuing spatially periodic responses are detected by diffraction of an optical probe beam. The use of polarization-selective detection allows for the isolation of magnetic and nonmagnetic responses. We observe oscillations of the diffracted signal intensity indicating magnetization precession due to both spin waves and magnetoelastic waves at the TG wave vector. We infer that the XTG excitation of magnetization precession is mediated by thermoelastically induced strain. We perform an analysis with the Landau-Lifshitz-Gilbert (LLG) equation that leads to an analytical solution closely reproducing the experimentally observed waveforms. We discuss the implications of our findings and further prospects for driving magnetic dynamics at the nanoscale using ultrashort x-ray pulses.

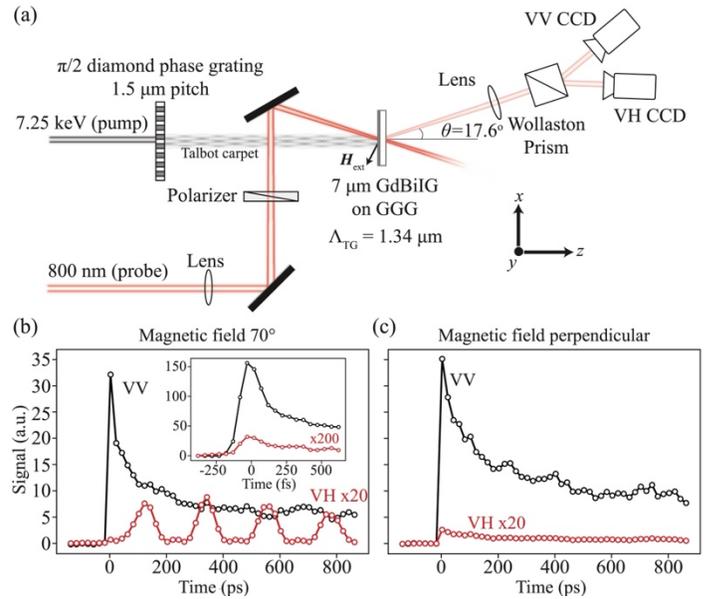

**Figure 1.** Experimental overview. (a) Top view of x-ray transient grating experimental setup. A spatially periodic x-ray intensity pattern in the sample is formed by Talbot imaging of a phase grating. An optical probe pulse is incident on the sample at the Bragg angle, and the diffracted signal is collected by a lens and passed through a Wollaston prism, which separates orthogonal polarizations. Two individual detectors simultaneously measure the VV (electronic and lattice) and VH (magnetic) responses. A permanent magnet placed on top of the sample generates magnetic field $H_{\text{ext}}$. (b) VV and VH responses at a magnetic field of 220 mT tilted at 70° to the sample normal. Inset shows the initial dynamics measured with fine time resolution. The VH signal at these early times is likely a leak-through from the VV channel due to imperfect alignment. (c) Responses measured with magnetic field of 220 mT normal to the sample surface. The signal is plotted on the same scale as in (b). Small VH responses in (b) and (c) are multiplied by factors indicated on the graphs.

The XTG experiments were performed at the Femtosecond X-ray Experiments beamline at the European XFEL [19]. A diagram of the experimental setup is shown in Fig. 1(a). The beamline providing ~100 fs instrument response function has been previously described in detail [17]. In this setup, monochromatized 7.25 keV pump pulses (~50 fs pulse duration, 0.8 eV FWHM, 0.6 µJ pulse energy at the sample, 70.5 kHz intra-train repetition rate) are diffracted by a diamond binary $\pi/2$-phase grating with a pitch of $p = 1.5$ µm placed 120 mm before the sample. For a plane incident wave, interference between the diffracted beams would lead to the formation of a Talbot carpet with planes of maximum contrast at odd multiples of $p^2/2\lambda_{\text{x-ray}} \approx 6.6$ mm along the propagation direction [18]. Due to a weak focusing of the x-rays, the separation of the Talbot planes changes slightly along the beam axis and leads to a smaller TG period at the sample of $\Lambda = 1.34$ µm, as determined by imprinting a permanent grating on a test sample (see Appendix A). This corresponds to a transferred in-plane wave vector of $k_{\text{TG}} = 2\pi/\Lambda = 4.7$ µm$^{-1}$. A vertically polarized, 800 nm, time-delayed probe pulse (~50 fs) is incident on the sample at the Bragg angle of $\theta = 17.6°$, defined by the phase matching condition. The spot sizes (full-width half-max) are about 100 µm for the pump and 50 µm for the probe.

The sample is a 7-µm-thick $(GdTmPrBi)_3(FeGa)_5O_{12}$ single crystal, from here on named GdBiIG, grown by liquid-phase epitaxy on a (111)-oriented gadolinium gallium garnet $(Gd_3Ga_5O_{12})$ [19]. Tm and Pr are low-concentration dopants added to adjust magnetic anisotropy, and Ga is added to enhance optical transparency. The GdBiIG thin film is a ferrimagnet with perpendicular magnetic anisotropy, a compensation temperature of ~240 K, and a Curie temperature of ~440 K. The external magnetic field vector, $H_{ext}$, lies in the scattering plane, and is generated by a permanent neodymium magnet. The angle and magnitude of the field vector are controlled by rotating the magnet placed just millimeters above the sample and changing its distance from the probe spot as shown in the Supplementary Material, S5 [20].

The diffracted beam propagates at a symmetric angle of $\theta_{out} = 17.6°$. While a Talbot image of a binary grating generally contains multiple Fourier components, our scattering geometry selects a single spatial Fourier component of the material excitations with a wave vector of $4.7~\mu m^{-1}$. The diffracted signal is collected by a lens and sent through a Wollaston prism, which directs orthogonal polarizations to separate charge-coupled-device (CCD) cameras. We denote the channel without polarization rotation as VV (vertical incident – vertical diffracted), and the channel with polarization rotation as VH (vertical incident – horizontal diffracted). While the detection setup superficially looks similar to that used to measure Faraday rotation, the idea here is different, as we are measuring the diffracted rather than the transmitted beam. Because periodic magnetization acts as a grating of dichroism, the signal diffracted by the periodic magnetization will have a 90° polarization rotation with respect to the incident probe beam [7] and can thus be isolated in the VH channel, while the modulation of the complex refractive index due to electronic and lattice responses will result in diffracted signal in the VV channel. Thus, the Wollaston prism separates the diffraction signals of magnetic and non-magnetic origins rather than measuring polarization rotation. Additionally, we need to compensate for the Faraday rotation caused by static magnetization by adjusting a wire-grid polarizer placed before the sample, as shown in the Supplementary Material, S4 [20].

Figure 1(b) shows an example of the response obtained for an external magnetic field canted at 70° to the sample normal. The VV signal shows a sharp peak at $t = 0$ which we ascribe to electronic excitation, followed by a decay that includes femtosecond and slower components. We interpret this behavior as the rapid decay of the initial electronic excitation followed by the slower decay of the thermal grating (the spatial modulation of the lattice temperature) via thermal diffusion. The VH signal, on the other hand, shows pronounced oscillations. The oscillations vanish when the magnetic field is normal to the sample, as shown in Fig. 1(c), confirming their magnetic origin and suggesting that their excitation is only possible under a canted external magnetic field. Subpicosecond dynamics are shown in the inset of Fig. 1(b). The ultrafast response in the VH channel is very small and identical in shape to the VV signal. We believe that this is likely a leak-through from the VV channel caused by imperfect rejection of the V polarization. The data provide no clear evidence of ultrafast demagnetization, which would be unlikely to yield a signal identical in shape to the electronic response in the VV channel, as the dynamics of electronic and magnetic responses should differ noticeably.

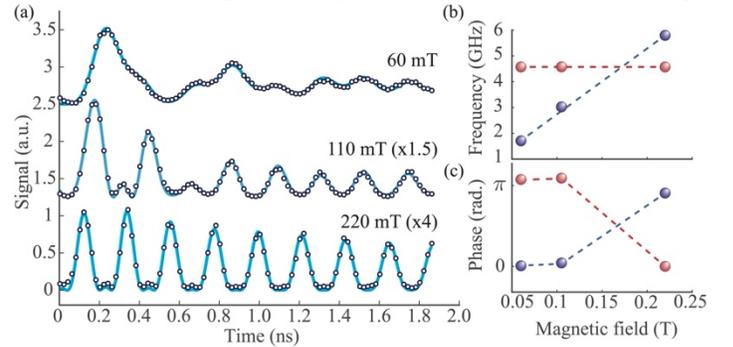

**Figure 2**. Magnetic XTG response. (a) VH response (open circles) measured at different magnetic field strengths with the direction of the magnetic field canted at 70° from the sample normal. Solid lines are two-frequency fits with Eq. 1. For clarity, experimental data and fits are normalized by the maximum amplitude of the experimental data at 60 mT and then rescaled by the factors shown in the plot. (b) Frequencies and (c) phases of the two modes extracted from the two-frequency model as a function of magnetic field.

Figure 2(a) shows the VH response at three representative magnetic field strengths with the canted orientation. The response is strongly dependent on the strength of the external field. The VV channel response did not show any dependence on the external magnetic field (see Supplementary Material, S2 [20]). All three waveforms are well described by a simple model consisting of two damped oscillatory terms and a constant offset:

$$S_{VH}(t) = \Theta(t)\left(a_1 e^{-\frac{t}{\tau_1}}\cos(\omega_1 t - \phi_1) + a_2 e^{-\frac{t}{\tau_2}}\cos(\omega_2 t - \phi_2) + C\right)^2 \quad (1)$$

where $\Theta(t)$ is the Heaviside step function. Here, the square is taken because the diffraction intensity from a magnetization TG is proportional to the square of the magnetization perturbation in the sample [7]. By fitting Eq. (1) to the data (solid curves in Fig. 2(a)), we extract the frequencies and phases of the two oscillatory terms [20]. Note that the presence of the constant term $C$ helps determine the absolute phases of the oscillations. As can be seen in Fig 2(b), one frequency varies linearly with the external magnetic field, while the other frequency remains constant at 4.6 GHz. There is a crossing point around 0.17 T, and the phases of the two modes exhibit a $\pi$-shift upon this crossing, as shown in Fig. 2(c) and also clearly visible in the time-domain data in Fig 2(a). The amplitude of the field-dependent mode decreases as the magnetic field strength increases; at 220 mT this mode is only visible as a small amplitude modulation of the constant-frequency mode. (This modulation is reproducible, as are the parameters of the field-dependent mode returned by the fit, see Supplementary Material, S1 [20].) We believe that the 4.6 GHz frequency corresponds to a magnetoelastic wave, i.e. a longitudinal acoustic wave accompanied by spin precession driven by inverse magnetostriction. The velocity $v = \omega/k_{TG}$ is 6.1 km/s, which is within the range of longitudinal acoustic velocities for garnets [21]. The measured velocity is slightly lower than that of yttrium-iron garnet (YIG) [21], which is unsurprising considering that Y is substituted with more massive Gd and Bi. The field-dependent frequency, on the other hand, is close to the expected ferromagnetic resonance frequency $\sim \gamma_0 H_{eff}$, where $\gamma_0$ is the standard gyromagnetic ratio of ~28 GHz/T, and $H_{eff}$ is the effective magnetic field that we assume to be dominated by the external field $H_{ext}$. While the diffraction signal originates from spin waves with a wave vector equal to the $k_{TG}$ rather than the homogeneous ferromagnetic resonance (FMR), at $k_{TG} = 4.7$ μm$^{-1}$, spin-wave dispersion is small so the spin wave frequency is close to that of the FMR [22].

Spin precession in ferromagnetic metal films driven by laser-generated surface acoustic waves via inverse magnetostriction has been previously studied in optical TG experiments [23]. Since in our case the film thickness and the x-ray penetration absorption length ($L \approx 6$ μm) are larger than the XTG period, we expect to excite longitudinal acoustic waves as in TG experiments performed in bulk materials [24]; hence, precession at the longitudinal acoustic frequency is not unexpected. Furthermore, we find that all three waveforms are very well reproduced by a simple model in which the magnetization precession is driven by a transient strain field induced by the XTG excitation. We assume that the excitation results in a spatially periodic temperature perturbation $\Delta T(t) = \Theta(t)\Delta T_{\max}(1 + \cos k_{TG}x)$. Based on the heat capacity of similar garnets [25], and the fluence and penetration depth of the XTG pump, we estimate this temperature increase to be $\Delta T_{\max} \approx$ 13 K at the TG maxima. This temperature perturbation gives rise to a strain field consisting of a quasi-stationary component following the temperature profile and slowly decaying via thermal diffusion, referred to as a "thermal grating," and counter-propagating longitudinal acoustic waves at wave vector $k_{TG}$ [26],

$$\epsilon_{xx}(t,x) = \Theta(t)\left(e^{-\frac{t}{\tau_{\text{therm}}}} - e^{-\frac{t}{\tau_{ac}}}\cos\omega_{ac}t\right) * \frac{1+\nu}{1-\nu}\beta\,\Delta T_{\max}(1+\cos k_{\text{TG}}x), \quad (2)$$

Where $\epsilon_{xx}$ is a strain tensor component, $\nu$ is Poisson's ratio, $\beta$ is linear thermal expansion coefficient, $\tau_{\text{therm}}$ is the thermal grating decay time, $\tau_{ac}$ is the acoustic attenuation time, and $\omega_{ac}$ is the acoustic wave frequency. The very weak elastic anisotropy of garnets such as YIG [27] justifies the use of the equation derived for elastically isotropic materials. The strain field, via the inverse magnetostriction effect, leads to a magnetoelastic contribution to the effective magnetic field,

$$\mu_0 \mathbf{h}(t,x) = \hat{x} 2b_1 \epsilon_{xx}(t,x) M_x, \quad (3)$$

where $b_1$ is the magnetoelastic coefficient. Eq. (3) shows that only the in-plane component, $M_x$, causes the magnetoelastic effect; hence an external magnetic field perpendicular to the sample surface will not lead to precession, in agreement with the experimental result shown in Fig. 1. In Appendix B, we solve the linearized Landau-Lifshitz-Gilbert equation [22] with the driving field given by Eq. (3) and arrive at the following analytical expression for the XTG magnetic signal,

$$S(t) \propto$$
$$\Theta(t)\left(c_1 e^{-\frac{t}{\tau_{\text{therm}}}} - c_2 e^{i\omega_{ac}t} - c_3 e^{-\frac{\gamma^2 H_0 \eta M_0}{1+\gamma^2\eta^2 M_0^2}t} e^{i\frac{\gamma H_0}{1+\gamma^2\eta^2 M_0^2}t} + \text{c.c}\right)^2 \quad (4)$$

where and $H_0$, $\gamma$, and $M_0$ are the external magnetic field strength, effective gyromagnetic ratio, and static magnetization, respectively, and the amplitude coefficients $c_i$ are given in Appendix B. The expression for $S(t)$ was fitted to the experimental data, using two free parameters, $\gamma H_0$ and $\gamma \eta M_0$, which can be interpreted as the natural magnon frequency and effective damping rate. The acoustic frequency, $\omega_{ac}/2\pi = 4.6$ GHz, was taken from the two-frequency model fit, and the thermal grating decay time, $\tau_{\text{therm}} \gg 1$ ns, was taken from the VV signal dynamics shown in supplemental Fig. S2 [20]. The resulting dynamics, shown as solid lines in Fig. 3(c), are in excellent agreement with the observed experimental results and fit nearly as well as the phenomenological two-mode model (Fig. 2(a)). In addition, we also performed measurements with a pump photon energy slightly above the Fe K-edge (7.135 keV), which showed identical dynamics apart from an amplitude factor (see Supplementary Material, S3 [20]), suggesting that the dynamics are not strongly wavelength-dependent as expected for the proposed mechanism.

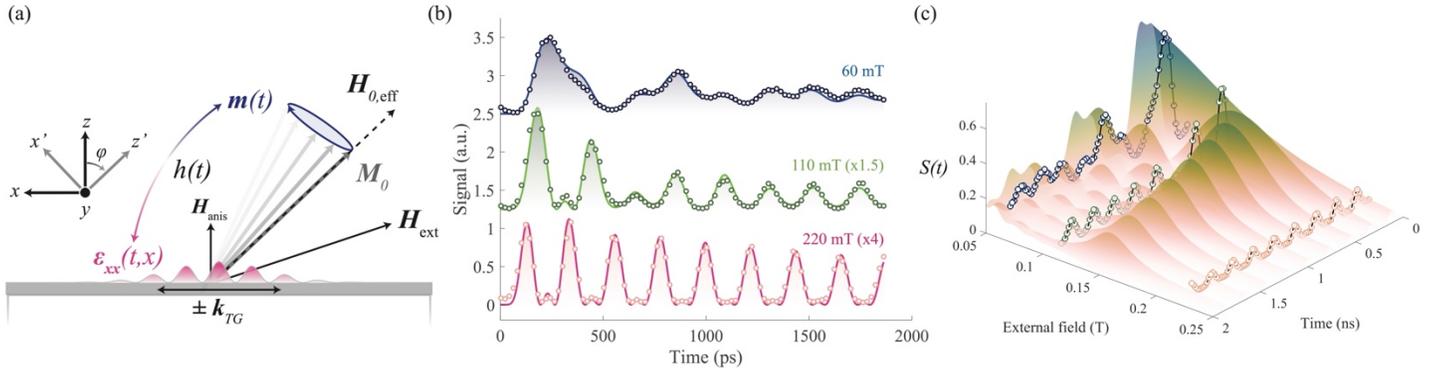

**Figure 3**. Micromagnetic model. (a) An external magnetic field $H_{ext}$ is applied at a canted angle, and the resulting initial effective magnetic field, $H_{0,eff}$ and equilibrium magnetization $M_0$ lie in between $H_{ext}$ and the anisotropy field $H_{anis}$ at an angle $\varphi$. We rotate our coordinate system around $y$ such that the equilibrium magnetization direction lies along $z'$. A strain field $e_{xx}(t,x)$ induced by the XTG excitation creates a time-varying magnetoelastic effective field $h(t)$ which drives magnetization dynamics, $m(t)$. (b) Analytical theory fits by Eq. (7) (solid curves) to the experimental data (open circles). The waveforms are rescaled to comparable amplitudes and vertically displaced for clarity. (c) 3D surface plot of a simulation done with Eq. (7) for a range of external magnetic fields. Experimental data points are overlayed as open circles.

A simulation for varying magnetic field obtained with the analytical solution is shown in Fig. 3(c) using the average gyromagnetic ratio and damping parameters extracted from the experimental data. The analytical model helps us understand the origin and behavior of the three terms making up the response. The oscillation at the acoustic frequency is the precession driven by the acoustic wave; essentially it is a classic damped harmonic oscillator response to a harmonic driving force, and a phase shift by $\pi$ upon crossing the resonance is expected. These oscillations are long lived, as their decay is given by an acoustic attenuation time greatly exceeding our experimental time window. However, our driving force is not a pure sinusoid: since $\mathbf{h}(t)$ is zero at $t<0$, it also has a step-function component, which adds spectral content at zero and nonzero frequencies. This gives rise to the second oscillatory term at the spin wave frequency $\omega_{FMR}$, essentially a response of a suddenly driven damped harmonic oscillator. The $\pi$ phase shift in this case is due to the behavior of the phase of the driving force, i.e. the right-hand side of Eq. (5). Finally, the non-oscillatory term is due to the quasi-static thermal expansion.

The fact that our model describes the measured waveforms very well indicates that there is no appreciable demagnetization, not only on the ultrafast time scale but within the entire 2 ns time window, and that the x-ray induced strain is the sole source of the observed magnetization dynamics. This behavior stands in contrast to what has been observed in ferro- and ferrimagnetic metal films both with optical and EUV excitation. For magnetic dielectrics, demagnetization on the sub-ps [25,28], ps [29], and ns [30] timescales has been reported.

It is somewhat surprising that the direct modulation of the refractive index by the standing acoustic wave does not result in acoustic oscillations in the VV channel. This can be rationalized if the refractive index of GdBiIG at the probe wavelength is much more sensitive to the electronic excitation and lattice temperature increase than to strain; as a result, the acoustic contribution in the VV signal is obscured. In the background-free XTG measurement, the noise is proportional to the signal itself (as one can see in Fig. 1(b), there is very little noise at negative time delays), therefore a large signal produced by e.g. a temperature modulation would make it more difficult to see the acoustic component of the VV signal.

To summarize, we have demonstrated TG measurements with hard x-ray excitation and optical probing, in which the signal is produced by x-ray driven magnetization precession in a magnetic dielectric. We observe spin waves and magnetoelastic waves at a wave vector of 4.7 μm$^{-1}$ and find that both responses are driven by x-ray generated strain via inverse magnetostriction. These results pave the way for using ultrashort x-ray pulses for driving magnetic dynamics. The XTG approach has a potential for exciting coherent and non-equilibrium magnetic dynamics across the entire Brillouin zone and detecting them with high frequency resolution enabled by time-domain detection. While in this study, the accessible wave vector range was limited by the use of an optical probe and did not exceed that of optical TG experiments [23,31], this limitation will be overcome in x-ray-pump / x-ray-probe TG experiments [32,33]. It is possible to produce x-ray interference patterns with periods below 1 nm using crystal optics [34], which can be combined with split and delay systems [35]. In addition to the transient grating technique, another way to reach large wave vectors is to exploit the localized nature of x-ray excitation by measuring diffuse scattering in a collinear x-ray pump-probe experiment as was done in a recent study [36]. Although no x-ray driven demagnetization was observed in these results, the recent studies in EUV TG measurements on itinerant magnetic systems [7,8,9,10] make us confident that these phenomena can also be driven by x-rays. Additionally, the long penetration depth of hard x-rays allows for pumping deep into bulk samples which is typically not the case for above-gap optical excitations. XTG may also be applied to studying spin diffusion [37]. High-wave-vector ultrafast studies of magnetic dynamics will make it possible to reach the fundamental limit where the spin mean free path exceeds the spatial period of the excitation and provide insights into nanoscale spin transport and ultrafast demagnetization, the behavior of coherent nm-wavelength spin waves and their coupling to other nanoscale material phases.

## Acknowledgements

C.S. and Ca.S. acknowledge financial support from the Spanish Ministry of Science, Innovation, and Universities through project PID2023-152154NB-C21. C.S. and W.G. acknowledge funding from the "Severo Ochoa" Programme for Centres of Excellence in R&D (CEX2020-001039S/AEI/10.13039/501100011033). P.R.M., A.A.M., N.B., and K.A.N received support from the Department of Energy, Office of Science, Office of Basic Energy Sciences, under Award Number DE-SC0019126. M.B, P.C. R.C, G.R. acknowledge support from Nanoscience Foundry and Fine Analysis (NFFA-MUR Italy Progetti Internazionali). R.C acknowledges the Italian Ministry of Foreign Affairs and International Cooperation (MAECI), Grant no. PGR12320 - U-DYNAMEC - CUP B53C23006060001. SC and MC acknowledge support of the ERC Advanced Grant CHIRAX (n˚ 101095012). D.R. is part of the Max Planck School of Photonics supported by the German Federal Ministry of Education and Research (BMBF), the Max Planck Society and the Fraunhofer Society. R.T. acknowledge the Next Generation EUProgramme: project PRIN-2022JWAF7Y [CUP: B53D23004250006] and I-PHOQS Infrastructure [IR0000016, ID D2B8D520, CUP B53C22001750006]. J.S. acknowledge Polish Ministry Science and Higher Education with projects 1/SOL/2021/2 and 2022/WK/13. W.G. acknowledges partial funding from Spanish Ministry of Universities through "Ayudas Beatriz Galindo" (BEAGAL18/00092), Regional Government of Madrid and Universidad Autónoma de Madrid through "Proyectos de I+D para Investigadores del Programa Beatriz Galindo" grant (Ref. SI2/PBG/2020-00003) and from Spanish Ministry of Science, Innovation and Universities through "Proyectos de I+D+i 2019" grant (Ref. PID2019-108678GB-I00) and "Proyectos de I+D+i 2022" grant (Ref. PID2022-140257NB-I00). The authors acknowledge the European XFEL facility for the Long Term Proposal 3323 granted to develop the X-ray transient grating technique. The authors would like to thank Alexander von Reppert for assistance in calibration of the external magnetic fields used in these measurements. In addition, the authors would like to thank Nina Rohringer for her insights and fruitful discussions.

## Data Availability

The raw data recorded for this experiment at the European XFEL are available [38].

# End Matter

*Appendix A: Imprinted grating*

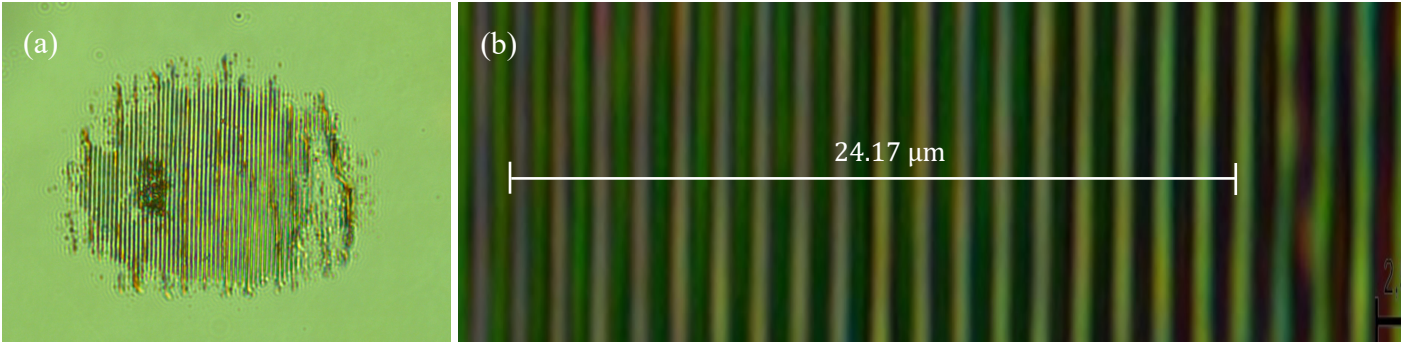

**Figure 4**. (a) Optical microscope image of an imprinted grating in $Bi_4Ge_3O_{12}$. (b) Zoomed in image used to measure the grating periodicity, which is found to be 1.34 ±0.01 μm.

For initial alignment purposes, and to measure the periodicity of the transient gratings formed by Talbot imaging of the 1.5 μm phase grating, permanent gratings were printed on bismuth germanate $Bi_4Ge_3O_{12}$ using an x-ray fluence above the damage threshold on a single shot basis. Optical microscope images (see Fig. 4) were taken and used to measure a periodicity of 1.34 μm for the imprinted gratings.

*Appendix B: Analytical micromagnetic model*

We describe the dynamics of magnetization at each spatial coordinate $x$ using the Landau-Lifshitz-Gilbert (LLG) equation of motion [22]:

$$\frac{\partial \mathbf{M}}{\partial t} = -\gamma \mathbf{M} \times \mathbf{H}_{\text{eff}} + \gamma \eta \mathbf{M} \times \frac{\partial \mathbf{M}}{\partial t}, \quad (A1)$$

where $\eta$ is the magnetic damping and $\mathbf{H}_{\text{eff}}(t)$ is the time-varying effective field. We disregard the dipole-dipole interaction in the spin wave, which is equivalent to neglecting the spin wave dispersion at our small wave vector. From Eqs. (2) and (3), we get an expression for the perturbation of the effective magnetic field,

$$h_x(t,x) = R(x)\Theta(t)\left(e^{-\frac{t}{\tau_{\text{therm}}}} - e^{-\frac{t}{\tau_{\text{ac}}}}\cos\omega_{\text{ac}}t\right), \quad (A2)$$

where $R(x) = \frac{2}{\mu_0} b_1 M_0 \sin\phi \frac{1+\nu}{1-\nu}\beta\, \Delta T(1 + \cos k_{\text{TG}}x)$.

Note that since we disregarded the dipole interaction, the spatial dependence of $R(x)$ does not affect the time dynamics. The time dependence of the driving field is given by

$$F(t) = \Theta(t)\left(e^{-\frac{t}{\tau_{\text{therm}}}} - e^{-\frac{t}{\tau_{\text{ac}}}}\cos\omega_{\text{ac}}t\right). \quad (A3)$$

To simplify the formulae, we rotate our coordinate system about the y-axis, $(x, y, z) \to (x', y, z')$, such that $z'$ is along the equilibrium magnetization direction as shown in Fig 3(a). We linearize the LLG by representing $\mathbf{H}_{\text{eff}}$ and $\mathbf{M}$ as sums of constant components directed along $z'$ and small time-dependent perturbations, $\mathbf{M} = \mathbf{M}_0 + \mathbf{m}(t)$, $\mathbf{H}_{\text{eff}} = \mathbf{H}_0 + \mathbf{h}(t)$.

Disregarding terms quadratic in the perturbation, we obtain a linearized LLG,

$$\frac{d\mathbf{m}(t)}{dt} = -\gamma \mathbf{M}_0 \times \mathbf{h}(t) - \gamma \mathbf{m}(t) \times \mathbf{H}_0 + \gamma \eta \mathbf{M}_0 \times \frac{d\mathbf{m}(t)}{dt}. \quad (A4)$$

The dynamical magnetization vector $\mathbf{m}(t)$ lies in the $x', y$-plane, but we are only interested in its $x'$-component, as our detection is not sensitive to $m_y$. Eq. (A7) yields two scalar equations for $x'$ and $y$ components of $d\mathbf{m}(t)/dt$, from which we get an equation for $m_{x'}$ in the form of a driven damped harmonic oscillator equation.

$$\frac{\partial^2 m_{x'}}{\partial t^2} + 2\zeta_0 \frac{\partial m_{x'}}{\partial t} + \omega_0^2 m_{x'}$$
$$= \frac{\gamma M_0}{1 + \gamma^2 \eta^2 M_0^2}\left(\gamma H_0 h_{x'} + \gamma M_0 \eta \frac{\partial h_{x'}}{\partial t}\right) \quad (A5)$$

where $\omega_0 = \frac{\gamma H_0}{\sqrt{1+\gamma^2\eta^2 M_0^2}}$ and $\zeta_0 = \frac{\gamma^2 H_0 \eta M_0}{1+\gamma^2\eta^2 M_0^2}$ are the frequency and damping rate of the oscillator, respectively. $H_0$ and $M_0$ are the magnitudes of vectors $\mathbf{H}_0$ and $\mathbf{M}_0$. We perform a Fourier transform of Eq. (A5) and obtain the following expression for the Fourier image of $m_{x'}(t)$,

$$m_{x'}(\omega) = \frac{CF(\omega) + i\omega DF(\omega)}{-\omega^2 + iA\omega + B}, \quad (A6)$$

where $F(\omega)$ is the Fourier transform of $F(t)$ obtained from Eq. (A3),

$$F(\omega) = \frac{1}{i\omega + \alpha_{\text{therm}}}$$
$$-\frac{1}{2}\left[\frac{1}{\alpha_{\text{ac}} + i\omega + i\omega_{\text{ac}}} + \frac{1}{\alpha_{\text{ac}} + i\omega - i\omega_{\text{ac}}}\right], \quad (A7)$$

and the terms A, B, C, D are given by

$$A = \frac{2\gamma^2 H_0 \eta M_0}{1 + \gamma^2 \eta^2 M_0^2} \quad (A8)$$

$$B = \frac{\gamma^2 H_0^2}{1 + \gamma^2 \eta^2 M_0^2} \quad (A9)$$

$$C = R(x) \frac{\gamma^2 H_0 M_0}{1 + \gamma^2 \eta^2 M_0^2} \quad (A10)$$

$$D = R(x) \frac{\gamma^2 M_0^2 \eta}{1 + \gamma^2 \eta^2 M_0^2} \quad (A11)$$

The time-domain solution is obtained via inverse Fourier transform:

$$m_{x'}(t) = \int_{-\infty}^{\infty} \frac{d\omega}{2\pi} e^{i\omega t} \frac{(C + i\omega D) F(\omega)}{-\omega^2 + iA\omega + B}, \quad (A12)$$

We define the closed contour integral over the complex variable $z$ in the upper half of the complex plane

$$m_{x'}(t) = \frac{1}{2\pi} \oint_C dz \, e^{izt} (C + izD) F(z) \frac{1}{-z^2 + iAz + B} \quad (A13)$$

The above integrand has 5 poles, all in the upper half of the complex plane:

$$z_1 = \omega_{ac} + i\alpha_{ac} \quad (A14)$$

$$z_1^* = -\omega_{ac} + i\alpha_{ac} \quad (A15)$$

$$z_2 = i\alpha_{therm} \quad (A16)$$

$$z_3 = \frac{iA}{2} + \frac{A}{2\gamma\eta M_0} \quad (A17)$$

$$z_3^* = \frac{iA}{2} - \frac{A}{2\gamma\eta M_0} \quad (A18)$$

The contour integral is given by the sum of the residues at the poles. Note that for each of the two pairs of symmetric poles we only need to compute the residue at the pole with the positive real part; the residue at the other pole will be given by the complex conjugate. Thus, we arrive at a final solution for the $x'$-component of magnetization

$$m_{x'}(x, t > 0) = R(x)\left(c_1 e^{-\frac{t}{\tau_{therm}}} - c_2 e^{i\omega_{ac} t} - c_3 e^{-\zeta_0 t} e^{i\omega_{FMR} t} + \text{c. c.}\right), \quad (A19)$$

where we assume the acoustic attenuation is negligible, $\alpha_{ac} = 0$, and define $\omega_{FMR} = \frac{\gamma H_0}{1 + \gamma^2 \eta^2 M_0^2}$. The coefficients $c_i$ in Eq. ($A19$) are given by the following expressions,

$$c_1 = \frac{\omega_{FMR} \gamma M_0 \left(1 - \alpha_{therm} \frac{M_0 \eta}{H_0}\right)}{\alpha_{therm}^2 - 2\alpha_{therm} \omega_{FMR} \gamma \eta M_0 + \omega_{FMR} \gamma H_0} \quad (A20)$$

$$c_2 = \frac{\omega_{FMR} \gamma M_0 \left(1 + i\omega_{ac} \frac{M_0 \eta}{H_0}\right)}{-2\omega_{ac}^2 + 4i\omega_{ac} \omega_{FMR} \gamma \eta M_0 + 2\omega_{FMR} \gamma H_0} \quad (A21)$$

$$c_3 = \omega_{FMR} \gamma M_0 \left(\frac{i}{2\omega_{FMR}} - \frac{i\gamma \eta^2 M_0^2}{2H_0} - \frac{M_0 \eta}{2H_0}\right) *$$

$$\left(-\frac{1}{\omega_{FMR}\gamma\eta M_0 - i\omega_{FMR} - \alpha_{therm}} \right.$$
$$+ \frac{1}{2\omega_{FMR}\gamma\eta M_0 - 2i\omega_{FMR} - 2i\omega_{ac}}$$
$$\left. + \frac{1}{2\omega_{FMR}\gamma\eta M_0 - 2i\omega_{FMR} + 2i\omega_{ac}}\right) \quad (A22)$$

The TG signal is proportional to the out-of-plane magnetization squared, i.e. $S(t) \propto |m_{x'}|^2$, which yields Eq. (4) in the main text.